\PassOptionsToPackage{dvipsnames}{xcolor}

\documentclass[sigconf]{acmart}
\usepackage{xspace}
\acmConference[EASE 2026]{The 30th International Conference on Evaluation and Assessment in Software Engineering}{Tue 9 - Fri 12 June 2026}{Glasgow, United Kingdom}

\setcopyright{none}

\newcommand{\Ourtoollong}{Hallucination Inspector\xspace}
\newcommand{\titlerunning}{\Ourtoollong\xspace}
\newcommand{\titlelong}{\Ourtoollong: A Fact-Checking Judge for API Migration\xspace}

\usepackage{listings}
\usepackage{xcolor}
\usepackage{algorithm}
\usepackage{algpseudocode}
\usepackage{tikz}

\usetikzlibrary{shapes, arrows, positioning, fit, calc}

\definecolor{cb-blue-sky}{RGB}{0, 110, 180}
\definecolor{cb-burgundy}{RGB}{145, 30, 60}
\definecolor{cb-green-sea}{RGB}{0, 150, 100}

\newboolean{showcomments}
\setboolean{showcomments}{false} 
\ifthenelse{\boolean{showcomments}}{
  \newcommand{\nbc}[3]{
    {\colorbox{#3}{\bfseries\sffamily\scriptsize\textcolor{white}{#1}}}%
    {\textcolor{#3}{\textsf\small$\blacktriangleright$\textit{#2}$\blacktriangleleft$}}}
  \newcommand{\todo}[1]{\nbc{TODO}{#1}{blue}\xspace}
}{
  \newcommand{\nbc}[3]{}
  \newcommand{\todo}[1]{}
}

\newcommand\javastyle{\lstset{
    language=Java,
    basicstyle=\small\ttfamily,
    otherkeywords={var}, 
    keywordstyle=\bfseries\color{cb-blue-sky},
    emph={setStreamType, USAGE_GAME_UNKNOWN}, 
    emphstyle=\bfseries\color{cb-burgundy}, 
    stringstyle=\color{cb-green-sea},
    showstringspaces=false,
    numbers=left,
    numberstyle=\tiny\color{gray},
    breaklines=true,
    frame=single,
    captionpos=b
}}

\lstnewenvironment{java}[1][]
{
    \javastyle{}
    \lstset{#1}
}
{}

\newcommand{\hinspector}{Hallucination Inspector\xspace}
\newcommand{\scahall}{Scaffolding Hallucination\xspace}
\newcommand{\phantoms}{Phantom Symbol\xspace}
\newcommand{\atomic}{Atomic Hallucinations\xspace}
\newcommand{\scope}{Scope-bound Hallucinations\xspace}

\begin{document}
\title{\titlelong}

\author{Marcos Tileria}
\email{m.tileriapalacios@surrey.ac.uk}
\affiliation{%
\institution{University of Surrey}
\city{Guildford}
\country{United Kingdom}
}

\author{Santanu Kumar Dash}
\email{s.k.dash@surrey.ac.uk}
\affiliation{%
\institution{University of Surrey}
\city{Guildford}
\country{United Kingdom}
}

\author{Profir-Petru P\^arțachi}
\email{partachi@comp.isct.ac.jp}
\affiliation{%
\institution{Institute of Science Tokyo}
\city{Tokyo}
\country{Japan}
}

\author{Earl T. Barr}
\email{e.barr@ucl.ac.uk}
\affiliation{%
\institution{University College London}
\city{London}
\country{United Kingdom}
}


\begin{abstract}
    Large Language Models (LLMs) are increasingly deployed in automated software engineering for tasks such as API migration. While LLMs are able to identify migration patterns, they often make mistakes and fail to produce correct glue code to invoke the new API in place of the old one. 
We call this issue \scahall, a failure mode where models generate incorrect calling contexts by inventing Phantom Symbols—such as imaginary imports, constructors, and constants—that do not exist in the API specification.
In this paper, we show that standard metrics cannot be relied upon to detect these instances of hallucination. We propose \hinspector, a static analysis tool to detect
\scahall in LLM-generated code. Our approach includes a lightweight evaluation framework that verifies symbols extracted from the abstract syntax tree against a knowledge base derived directly from software documentation for the API.
A preliminary evaluation on Android API migrations demonstrates that our approach successfully identifies hallucinations and significantly reduces false positives compared to standard metrics and probabilistic judges.

\end{abstract}

\maketitle
\title{\titlerunning}


\section{Introduction}
\label{sec:intro}

The advent of Large Language Models (LLMs) has fundamentally altered the landscape of automated software engineering, offering unprecedented capabilities in code generation, translation, and repair. 
This potential has driven the rapid adoption and integration of LLMs into IDEs and CI/CD pipelines to automate tedious software maintenance tasks such as API migration and dependency updates~\cite{11121699,maddila2025agentic,10.1145/3696630.3728542}.
Despite their capabilities, LLMs often generate code with errors such as hallucinations~\cite{zhang2025llm,pan2024lost,twist2025library} that are often difficult to detect.

Within the context of API migration, which is the focus of our work, we refer to the old API as the \emph{source} and the new API as the \emph{target} of the migration. LLM generation involves two steps. 
In the first step, the LLM identifies the target API to use. In the second step, it generates the glue code required to construct objects and invoke the target API, including field accesses, builder calls, and method chains, as well as any additional code needed to use the API outputs. 
Hallucinations at any stage of the process will result in incorrect migrations. Therefore, reliably detecting them is essential to reduce false positives in the migration process.

Hallucinations have often been used loosely to describe any generation error. In this paper, we adopt a strict definition tailored for this domain: hallucination is the generation of content that contradicts verifiable facts contained within a reference knowledge base.
In the specific domain of API migration, this knowledge base is the API specification described in its documentation. Therefore, a hallucination occurs specifically when an LLM generates a symbol that does not exist in the API documentation.

While recent studies show that LLMs can correctly identify the target~\cite{11329192,wang2024llms}, they struggle to generate glue code required for valid instantiation~\cite{10549604,paudel2025hallucinot,zhang2025llm} and usage of return values from the API call. 
To formalise this observation, we introduce the novel concept of \emph{\scahall}: a failure mode where the model hallucinates the code required to build the context for calling the target API and use its results.
We refer to these invented tokens---imports, API names, methods, or constants not present in the API's documentation---as \emph{\phantoms}.

In this paper, we show that detecting \scahall is challenging. Generated code is often quite similar to the correct solution which defeats both automated and manual approaches.
Standard similarity metrics like CodeBLEU~\cite{ren2020codebleu} are used to measure textual and syntactic overlap with a ground truth. Due to the similarity of the generated code with the correct solution, a hallucinated patch often receives a high CodeBLEU score. Even a manual approach to evaluation is prone to such errors and it is prohibitively expensive for large scale migrations. 
Recently, techniques have been proposed to scale the evaluation using ``LLM-as-a-Judge'', where an LLM is used to check correctness of the migration. However, these have been found to be unreliable, showing both optimism bias~\cite{10.1145/3728963,10549604} and over-correction bias~\cite{ASE63991.2025.00323}.

To address the limitations of existing approaches, we propose a novel solution for detecting \scahall which we call The \hinspector. It is a knowledge-based fact-checking evaluation framework designed for detecting \phantoms in LLM-generated code. 
Similar to previous work that use API specifications to guide software engineering tasks~\cite{11121691,tileria2024docflow}, our knowledge base is derived directly from the API reference. Therefore, our inspector serves as a evaluator grounded in an API oracle unlike similarity metrics or LLM judges.
We parse the generated patch into an Abstract Syntax Tree (AST) and cross-reference the extracted symbols against our knowledge base that is derived from a trusted specification. Therefore, we can verify hallucinations of method calls, constants, and type instantiation. 
This approach allows us to move beyond simple plausibility checks and proactively flags broken migrations that superficially resemble correct code but contain \scahall.

Our contributions are as follows:
\begin{itemize}
    \item A definition of \scahall grounded in the contradiction of a knowledge base covering two distinct categories --- \emph{Atomic} and \emph{Scope-bound} Hallucinations.
    \item The \hinspector, a static analysis tool to detect \scahall in LLM-generated code. 
    \item A preliminary evaluation on a dataset of Android API migrations, demonstrating that \hinspector improves LLM-as-a-Judge baseline and that standard metrics like CodeBLEU are not reliable for detecting \scahall.
\end{itemize}

The source code for the \hinspector and the artefacts are available in our repository\footnote{\url{https://zenodo.org/records/19633303}}.

\section{Background and Motivation}
\label{sec:motivation}

Automated API migration faces a fundamental API fidelity problem. Generated code must capture the correct intent while remaining strictly compliant with the calling protocol for the target API or library.
While LLMs often correctly identify the target replacement, they struggle with details that hold the migration together. These details often involve builders, constants, call chains, etc. 
We use the term \scahall to describe the phenomenon where LLMs hallucinate API attributes that do not exist in the software documentation.

Listing~\ref{lst:hallucination} shows the migration of a legacy \lstinline|MediaPlayer| API call in Android. The deprecated method \lstinline|setAudioStreamType| requires a simple integer constant. 
The replacement, \lstinline|setAudioAttributes|, requires constructing an \lstinline|AudioAttributes| object using a Builder pattern. The LLM generates a patch that looks plausible but contains instances of \scahall, as it fails to verify API usage against the actual API specification.

The snippet illustrates two distinct failure modes. First, line~\ref{line:hallucination1} invokes \lstinline|setStreamType| on the wrong object \lstinline|Builder|. Second, line~\ref{line:hallucination2} uses \lstinline|CONTENT_TYPE_STREAM|, which does not exist. 
Correct usage would require calling \lstinline|setUsage| with a valid constant such as \lstinline|CONTENT_TYPE_MUSIC| and removing the invalid \lstinline|setStreamType| call, which is a method of the legacy \lstinline|AudioManager| class, not the new \lstinline|AudioAttributes.Builder|.
Both errors would cause compilation failures, yet standard evaluation metrics easily overlook them. 

\subsection{A Taxonomy of Verifiable Hallucinations}
\label{sec:taxonomy}

Previous work relies on high-level classifications such as API Conflicts~\cite{zhang2025llm} and API Mismatches~\cite{pan2024lost} to denote API hallucinations. 
However, we found these labels too broad to be actionable for automated verification. To address this, we systematically broke down these cases based on the structural complexity required to detect them, resulting in two distinct and actionable categories.
This distinction drives the architecture of our \hinspector, which employs different strategies for each category.

\noindent\textbf{Tier 1: \atomic.} This category refers to the invention of atomic symbols, e.g., classes, constants, or static methods that do not exist in the API namespace. We term these \emph{Phantom Symbols}. In Listing~\ref{lst:hallucination}, \lstinline|CONTENT_TYPE_STREAM| is a phantom constant. 
A hallucination checker like \hinspector that has access to the documentation for \lstinline|AudioAttributes| can easily detect \emph{Phantom Symbols}.

 \noindent\textbf{Tier 2: \scope.} The second category of hallucinations that we cover relate to fully-qualified symbols. Often, LLMs generate code in which they hallucinate methods for an object. In our example, the method \lstinline|setStreamType| exists in the legacy \lstinline|AudioManager| class but not in \lstinline|AudioAttributes.Builder|. To detect this, the inspector must traverse the call chain and resolve the variable to determine that the object is of type \lstinline|Builder|, not \lstinline|AudioManager|. We perform this resolution in \hinspector, and then consult API class documentation to detect these errors. \scope include (1) Misplaced Members (valid methods used on the wrong object) and (2) Invented Members (completely non-existent methods invoked within a chain). We term these \emph{Phantom Members}. 


\begin{figure}[t]
\begin{java}[caption={Deprecated API migration example. LLM-generated patch to migrate setAudioStreamType to the new version setAudioAttributes(AudioAttributes). The LLM hallucinates glue code to generate the AudioAttributes object.}, label=lst:hallucination, escapeinside={(*@}{@*)}]
// Legacy Code:
mediaPlayer.setAudioStreamType(AudioManager.STREAM_MUSIC);

// Generated Patch:
AudioAttribute attr = new AudioAttribute.Builder()
    .setStreamType(AudioManager.STREAM_MUSIC)  (*@\label{line:hallucination1}@*)
    .setContentType(AudioAttributes.CONTENT_TYPE_STREAM) (*@\label{line:hallucination2}@*)
    .build();
mediaPlayer.setAudioAttributes(attr);
\end{java}
\vspace{-1em}
\end{figure}

\section{The Approach: The Fact-Checking Judge}
\label{sec:approach}

The Hallucination Inspector is designed as a lightweight, documentation guided, static analysis framework. The architecture is language-agnostic and relies on two core principles:
Firstly, we employ a static checking strategy. Unlike dynamic approaches that require a buildable environment and test suite execution, our system operates directly on partial code snippets, allowing for rapid feedback without the overhead of compiling entire projects. 
Secondly, we make the detection deterministic by using an API oracle which reduces \emph{sycophancy} and \emph{optimism bias} inherent in LLM-based judges; the API oracle is derived from official documentation, rather than another probabilistic model. 

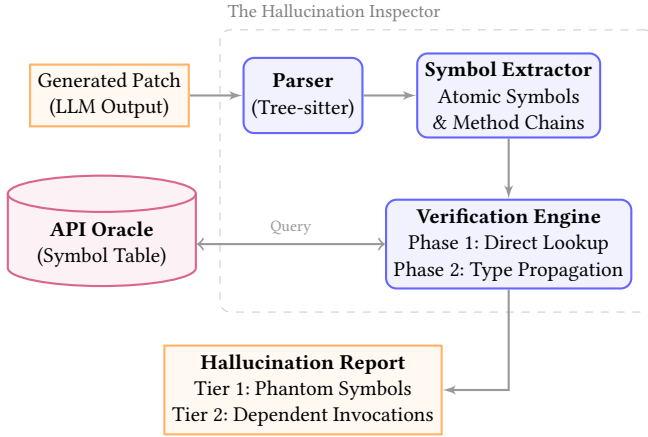
\begin{figure}[tbp]
\centering
\begin{tikzpicture}[
    node distance=0.5cm and 0.7cm,
    auto,
    process/.style={rectangle, draw=blue!60, fill=blue!5, thick, rounded corners, minimum height=1cm, align=center, font=\small},
    data/.style={rectangle, draw=orange!60, fill=orange!5, thick, minimum height=0.8cm, align=center, font=\small},
    oracle/.style={cylinder, shape border rotate=90, draw=purple!60, fill=purple!5, thick, aspect=0.25, minimum height=1.2cm, minimum width=2.5cm, align=center, font=\small},
    group/.style={rectangle, draw=gray!40, dashed, inner sep=0.3cm, rounded corners},
    line/.style={draw, -latex', thick, gray!80}
]

    \node [data] (input) {Generated Patch\\(LLM Output)};
    
    \node [process, right=of input] (parser) {\textbf{Parser}\\(Tree-sitter)};
    
    \node [process, right=of parser] (extractor) {\textbf{Symbol Extractor}\\ {Atomic Symbols} \\ \& {Method Chains}};
    
    \node [process, below=of extractor, yshift=-0.3cm] (verifier) {\textbf{Verification Engine}\\Phase 1: Direct Lookup\\Phase 2: Type Propagation};
    
    \node [oracle, left=2.5cm of verifier] (db) {\textbf{API Oracle}\\(Symbol Table)};
    
    \node [data, below=of parser, yshift=-2.3cm] (output) {\textbf{Hallucination Report}\\Tier 1: Phantom Symbols\\Tier 2: Dependent Invocations};

    \node [group, fit=(parser) (extractor) (verifier)] (core) {};
    \node [above right, font=\footnotesize\color{gray}] at (core.north west) {The Hallucination Inspector};

    \path [line] (input) -- (parser);
    \path [line] (parser) -- (extractor);
    \path [line] (extractor) -- (verifier);
    \path [line] (verifier) |- (output.east);
    
    \path [line, <->] (verifier) -- node[above, font=\scriptsize] {Query} (db);

\end{tikzpicture}
\caption{The Architecture of the Hallucination Inspector. The system parses generated code into an AST, extracts testable units, and verifies them against the API Oracle using a two-phase static analysis strategy.}
\label{fig:architecture}
\end{figure}

The pipeline consists of three sequential modules: a parser that converts code tokens into an AST; a symbol extractor that isolates API usages such as method calls and constants; and the API oracle that cross-references these symbols against a trusted specification, which is the API documentation in our case. Figure~\ref{fig:architecture} illustrates our workflow. Our lightweight architecture aligns with recent findings from industrial deployments, which argue that parsing tools paired with static checking are preferred over complex, latency-heavy LLM agent loops in large-scale codebase maintenance~\cite{Li2025BitsAIFixLA}.

\subsection{The Deterministic Oracle}
To enable efficient checking, we construct a ground-truth Symbol Table derived directly from the official Android API specification shipped with the SDK in XML format. Unlike raw source code, which often contains internal or hidden methods, this specification represents the strict public contract between library maintainers and developers, ensuring we only validate legally accessible symbols.
The processing pipeline first extracts classes, constants, and methods from JVM descriptors then recursively flattens the inheritance hierarchy into a hash table. 
For every class, the index stores a set of distinct method signatures (e.g., \texttt{setUsage(int)}), comprising both declared members and those inherited from superclasses. 
This pre-computation ensures that verifying a method call is an efficient set membership check eliminating the need for runtime traversal of the inheritance tree.

\subsection{Symbolic and Structural Verification}
The core of \hinspector is a static analysis engine that verifies the extracted AST symbols against the Oracle. 
The system parses the LLM-generated snippet to generate a concrete syntax tree. We traverse this tree to extract three categories of units: (1) Atomic Symbols, including static constants and simple field accesses; (2) Singular method calls; and (3) method chains, which is a composition of singular calls, representing fluent API usage.

The checking process is divided into two phases, targeting different classes of hallucination.
Phase 1 targets \atomic (e.g., phantom constants) via direct existence checks. Phase 2, detailed in Algorithm~\ref{alg:propagation}, handles \scope (e.g., phantom members in method chains) by tracking type propagation through local scope resolution.
Crucially, the process maps local variables to their underlying API types, ensuring the oracle only validates the existence of the types and invoked members.

The algorithm takes an invocation sequence $C$ and a local scope $S$. 
For method chains in $C$, we start from the left and resolve the first object  to determine its {Root Type}, which we call ($T_{curr}$). If there are additional calls, we resolve them incrementally by parsing the call chain from left to right. In a call chain of the form $f.g().h()$, we identify the type of $f$, to check valid membership, followed by a check of $f.g()$ if it is a valid call. If a member is missing in the API documentation, it is flagged as a {Phantom Member}. If valid, the algorithm updates $T_{curr}$ to the member's return type. If a method returns \texttt{void} but is followed by another call, it is flagged as a {Broken Chain}.

\begin{algorithm}[t]
\caption{Sequential Symbol Validation Algorithm}
\label{alg:propagation}
\begin{algorithmic}[1]
\Require Invocation Sequence $C = \{root, m_1, m_2, ..., m_n\}$, Scope $S$, Oracle $\mathcal{O}$
\Ensure Set of Hallucinations $H$

\State $H \gets \emptyset$

\If{$root$ is \textbf{StaticClass}}
    \State $T_{curr} \gets root$ 
\ElsIf{$root$ is \textbf{Variable}}
    \State $T_{curr} \gets S[root]$ 
\ElsIf{$root$ is \textbf{Constructor}}
    \State $T_{curr} \gets root.type$
\EndIf

\For{$i \gets 1$ \textbf{to} $n$}
    \State $m_i \gets C[i]$
    
    \If{$m_i \notin \mathcal{O}[T_{curr}]$}
        \State $H \gets H \cup \{\text{PhantomMember}(m_i, T_{curr})\}$
        \State \textbf{break}
    \EndIf
    
    \State $R_{type} \gets \mathcal{O}[T_{curr}][m_i].returnType$
    
    \If{$R_{type} == \text{void}$ \textbf{and} $i < n$}
        \State $H \gets H \cup \{\text{BrokenChain}(m_i)\}$
        \State \textbf{break}
    \EndIf
    
    \State $T_{curr} \gets R_{type}$ 
\EndFor

\State \Return $H$
\end{algorithmic}
\end{algorithm}

This dual-phase approach allows us to verify Atomic and Scope-bound Hallucinations where the LLM invents API symbols that do not exist in the target library, provided the instantiation occurs within the generated patch.
We implemented the \hinspector in Python using the TreeSitter library for parsing and a custom-built indexer for the Android API level $35$. The API documentation was sourced from the XML files provided by the Android SDK. 
The code is available in the supplementary materials.

\section{Experiments}
\label{sec:eval}

To validate the efficacy of \hinspector, we designed a two-part evaluation. Firstly, we analyse the limitations of standard similarity metrics, demonstrating why CodeBLEU is fundamentally unsuitable for detecting \scahall in API migrations. Next, we assess the detection performance of our tool against a LLM-as-a-judge baseline.  We address the following Research Questions:

\textbf{RQ1:The Blind Spot of metrics.} Do standard metrics fail to identify hallucinations? We use codeBLEU to measure the similarity of generated snippets to the ground truth, hypothesising that it will assign high similarity scores to incorrect API migrations.

\textbf{RQ2: Detection Efficacy.} Can the \hinspector accurately distinguish between valid context code and hallucinations? We report the Precision and Recall of our tool against a human-annotated ground truth, treating the oracle's flags as the predicted labels.
We compare our results against a LLM-as-a-Judge baseline, where a GPT-5 model is prompted to evaluate the generated patch and detect \scahall.

\subsection{Datasets}
\noindent\textbf{API Migration Dataset.} We constructed a dataset of 51 distinct API migration pairs. 
Each pair consists of a snippet with deprecated API usage, and its corresponding migration (ground truth). The snippets include necessary file-level context, such as imports and surrounding logic.
The pairs were sampled from previous work~\cite{haryono27androevolve,mahmud2024automated}, and the selection criterion prioritised migration scenarios where the new API method requires instantiating new types and adapting parameters rather than a simple rename.

\noindent\textbf{Automated Repair Task.} We tasked the LLMs with generating migration snippets while preserving functional equivalence. The prompt includes the code snippet with deprecated usage, along with the source and target APIs. As the target API method is explicitly provided, the LLM can focus on glue code to call the API. 
We selected two state-of-the-art open-weights models {Qwen-2.5-32b-Instruct} (Qwen-2.5) and {Codestral-v1-22b} (Codestral-v1) capable of running on our hardware, a server equipped with two NVIDIA RTX A5000 GPUs.

\noindent\textbf{Ground-Truth \& Taxonomy.} We established the ground-truth through manual annotation. Two authors inspected the LLM output and categorise the migration as valid based on correct migration and absence of hallucinations. 
The criteria for correct migration includes the correct use of API methods, valid syntax and semantics.
We categorise API failure modes into two classes:
(1) {API Misuse:} Using a valid but incorrect API element for the context such as the wrong constant or API method.
(2) {Hallucination:} Inventing a non-existent atomic and scope-bounded symbol. 
To evaluate the Inspector, we treat only hallucinations as the positive class to isolates verifiable hallucinations from general logic bugs. As shown in Listing~\ref{lst:hallucination}, a single migration may contain multiple hallucinations.

\subsection{Evaluation}
\label{sec:eval:results}

To establish a baseline for hallucinations, we first looked at the repair failure rate of the two models on our dataset of 51 Android API migrations. Table \ref{tab:taxonomy_breakdown} shows the distribution of errors. 
The Qwen2.5 model generated 32 invalid migrations, $26$ of which contain hallucinations , while Codestral failed on $36$ cases with $21$ hallucinations. 
Categorising these errors according to our taxonomy reveals that the models frequently fail due to both \atomic and \scope.
Within the snippets generated by Qwen2.5, we identified $15$ \atomic, which include phantom constants, imports, and static calls. We also found $28$ \scope, representing phantom instance calls and method chains evaluated against a base type. 
Similarly, the Codestral snippets contained $12$ \atomic and $25$ \scope. Because a single broken migration often contains multiple \phantoms, the total count of hallucinations exceeds the number of failed snippets. 
This distribution suggests that models are more susceptible to scope-bound errors where they hallucinate relationships between valid symbols and incorrect target types.

\subsubsection{RQ1: The Blind Spot of Similarity Metrics}
\label{sec:eval:metrics:blindspot}

\begin{figure}[t!]
    \centering
    \includegraphics[width=\columnwidth]{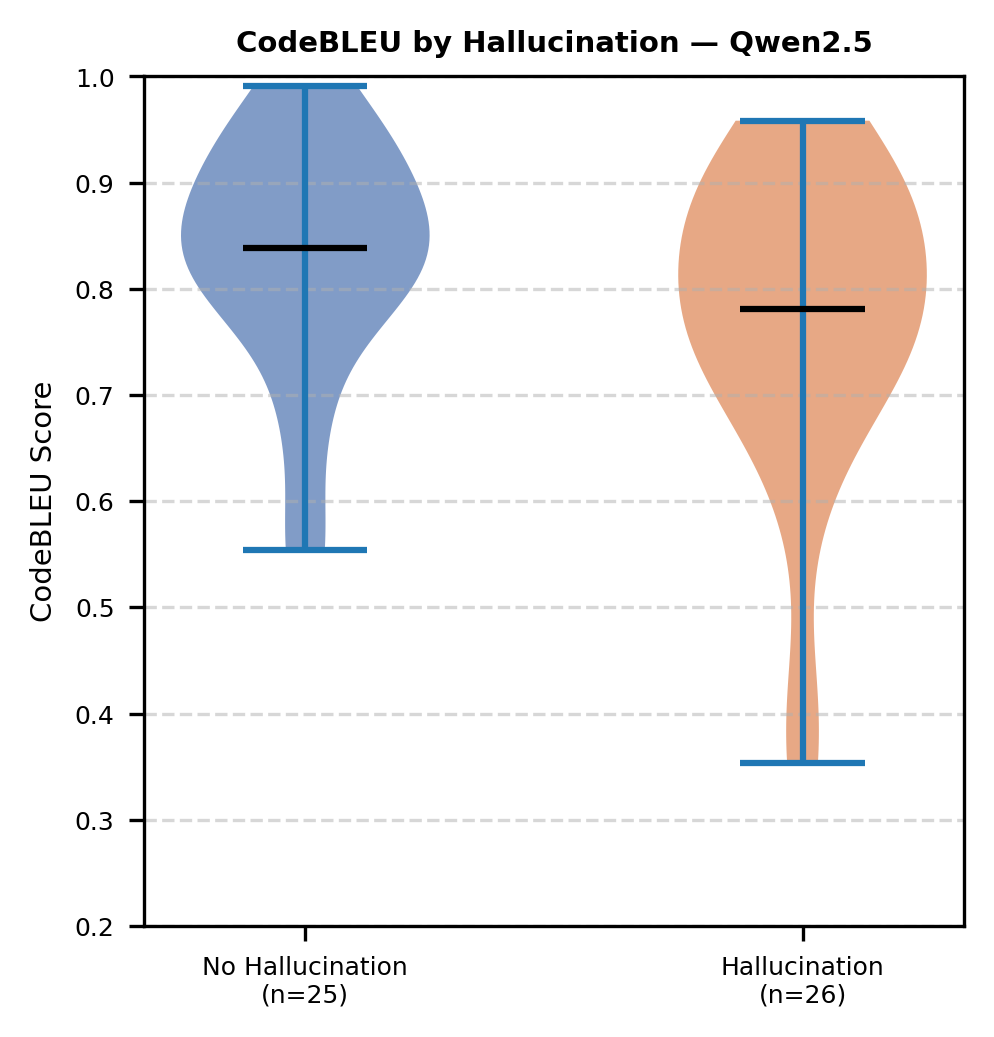}
    \caption{CodeBLEU score distribution for LLM migrations with and without Scaffolding Hallucinations.}
    \label{fig:codebleu_distribution}
\end{figure}

To evaluate the efficacy of standard similarity metrics in detecting \scahall, we analysed the CodeBLEU scores assigned to the generated snippets. 
The results demonstrate a fundamental limitation of relying on similarity metrics for API migration verification. A deeper look into CodeBLEU scores reveals a near total overlap between factually correct snippets and those containing hallucinations.
~\autoref{tab:codebleu_stats} shows that the median CodeBLEU score for hallucinations is comparable to the median for migrations without them, reducing reliabillity of CodeBLEU as a metric. Figure \ref{fig:codebleu_distribution} visually illustrates this overlap, showing that the presence of hallucinations does not significantly affect the CodeBLEU score. 
We show this for the Qwen2.5 model, which produces a similar number of patches with and without hallucinations. 
Despite this, codeBLEU and similar metrics are commonly used in the evaluation of code generation tasks~\cite{ren2020codebleu,amin2026jmigbench,wang2024llms,patil2025berkeley}.

\begin{table}[t!]
    \centering
    \caption{Migration Failures and Hallucination Taxonomy by Model. Dataset of 51 Android API migrations.}
    \label{tab:taxonomy_breakdown}
    \begin{tabular}{lcccc}
        \toprule
        Model & Invalid & Hallucinated & Atomic & Scope- \\
         & Migrations & Cases & & Bound \\
        \midrule
        Qwen2.5 & 32 & 26 & 15 & 28 \\
        Codestral & 36 & 21 & 12 & 25 \\
        \bottomrule
    \end{tabular}
\end{table}

\begin{table}[t!]
\centering
\caption{CodeBLEU score statistics by model. Compares the distribution of CodeBLEU scores for patches containing Hallucinations (Hallu.) against those without (No Hallu.)}
\label{tab:codebleu_stats}
\begin{tabular}{lcccc}
\toprule
& \multicolumn{2}{c}{Qwen2.5} & \multicolumn{2}{c}{Codestral-v1} \\
\cmidrule(lr){2-3} \cmidrule(lr){4-5}
Metric & Hallu. & No Hallu. & Hallu. & No Hallu. \\
\midrule
Mean & 0.791 & 0.807  & 0.624 & 0.694 \\
Median & 0.844 & 0.827 & 0.705 & 0.762 \\
Minimum & 0.353 & 0.553 & 0.135 & 0.261 \\
Maximum & 0.958 & 0.991  & 0.902 & 0.897 \\
\bottomrule
\end{tabular}
\end{table}

We found that a snippet containing \scahall can achieve a similarity score of nearly 0.96. Consequently, it is infeasible to establish a score threshold that reliably filters out hallucinations without discarding valid code. 
These findings confirm that an LLM can generate a close replica with high similarity score with the ground truth that differs by a single \phantoms while being uncompilable. 
This validates the necessity of our deterministic \hinspector, which successfully flags these hidden errors.

\subsubsection{RQ2: \hinspector vs LLM-as-a-Judge}
\label{sec:eval:efficacy}

To establish an automated evaluation baseline, we compared a gpt-5-mini LLM judge against our deterministic \hinspector. 
We selected gpt-5-mini for its efficiency and superior performance over the open-weights generation models. The judge was tasked with identifying \scahall and explain its reasoning. 
The prompt included the source and target API signatures, the original legacy snippet, and the candidate migration patch. Full details of the prompt are available in our supplementary artefact.
Under our strict evaluation criteria, a detection was only valid if both the positive label and the underlying explanation matched our manual annotations. 
Ultimately, the results in table~\ref{tab:llm_judge_summary} reveal severe reliability issues: the LLM judge failed in both directions, frequently approving broken code and rejecting valid patches.

\begin{table}[t]
\centering
\caption{Detection Performance Metrics for our approach (Hall. Inspector) and LLM-as-a-Judge Evaluators.}
\label{tab:llm_judge_summary}
\begin{tabular}{llccc}
\toprule
Evaluator & Generator & Precision & Recall & F1 \\
\midrule
Hall. Inspector & Qwen2.5 & 1.00 & 0.73 & 0.84 \\
Hall. Inspector & Codestral-v1 & 1.00 & 0.90 & 0.95 \\
gpt-5-mini & Qwen2.5 & 0.67 & 0.67 & 0.67 \\
gpt-5-mini & Codestral-v1 & 0.50 & 0.56 & 0.53 \\
\bottomrule
\end{tabular}
\end{table}

The Hallucination Inspector achieved perfect precision across both datasets. By grounding its verification in the API oracle rather than probabilistic inference, it completely eliminated false alarms.
The false negatives were primarily due to argument types and  symbols that coincidentally matched valid API elements, such as a non-existent method name that is identical to an existing one but with different parameters.
The LLM judge generated 8 false positives for Qwen2.5 and 10 for Codestral. When evaluating the Codestral patches, half of the instances flagged by the judge as hallucinations were actually valid API usages. This high rate of false positives demonstrates a severe over-correction bias, which would force an automated repair pipeline to discard correct implementations.
The probabilistic judge also failed to detect 8 actual hallucinations in both generator datasets. This poor recall indicates that the model frequently assumes that a superficially plausible API structure is factually correct, falling victim to the same probability divergence as the generator models.
Overall, the experimental data highlights that LLMs struggle with correct usage and validation of API usage. 

\section{Discussion and Future Work}
\label{sec:discussion}
\hinspector outperforms other metrics and achieves high precision in identifying hallucinations. False negatives in our evaluation stem from argument type mismatches and hallucinated callback signatures.
Our abstract type propagation covers name and return-type resolution, but lacks full signature resolution; we plan to extend it accordingly. We also found one case where a valid symbol was used incorrectly as a static method, and plan to augment the documentation processor to flag such modifier mismatches.

A key limitation is that we underperform for callbacks, since API migration must use them in a type-correct manner. We plan to extend the parser to validate callee-side declarations against callback documentation. LLM-generated code is also often incomplete, causing us to under-approximate type resolution; expanding the search scope deeper into the project would improve this.
Furthermore, our evaluation is constrained by a small dataset, a single LLM judge, and the Android Java ecosystem. Future work will expand to multi-language benchmarks and diverse evaluator models.

\hinspector can act as an agentic gatekeeper in an iterative repair loop, prompting the model to fix phantom symbols before final output.
Recent industrial frameworks, such as the ByteDance approach~\cite{Li2025BitsAIFixLA}, have successfully used static linting results as reward signals to train code repair models via Reinforcement Learning.
Similarly, \hinspector can reward RL pipelines and potentially train code generation models that are sensitive to scope-bound errors.

\section{Related Work}
\label{sec:related}

Automated API migration has evolved from rule-based tools~\cite{haryono27androevolve,yanjie2022towards,ramos2023melt,fazzini2020apimigrator} to LLM-based approaches, which introduce new reliability risks~\cite{mahmud2024automated,amin2026jmigbench,11121699,wang2024llms}.
Prior work has characterised the resulting failure modes: Zhang et al.\ taxonomised hallucinations into task-requirement and factual-knowledge conflicts~\cite{zhang2025llm}; Pan et al.\ identified API mismatch as a dominant failure in code translation~\cite{pan2024lost}; and Twist et al.\ detected name and member fabrications via regular expressions~\cite{twist2025library}.
Our work builds on these foundations by formalising \scahall into an actionable taxonomy that is checked through deep structural analysis rather than pattern matching.

To suppress hallucinations, RAG-based methods augment generation with documentation~\cite{11121691} or constrain decoding to library structures~\cite{10.1145/3696630.3728569}.
These approaches reduce error frequency but remain probabilistic; \hinspector instead focuses on post-generation verification, deterministically validating outputs against an API oracle.
Static evaluation of LLM-generated code has been approached via LLM-based detectors~\cite{10549604}, self-critique~\cite{dou2024s}, and LLM-as-judge frameworks~\cite{zheng2023judging,10.1145/3728963,ASE63991.2025.00323,moon2025don,koutcheme2025evaluating}, all of which exhibit optimism or over-correction bias.
By grounding assessment in documentation-derived specifications, \hinspector eliminates these biases and ensures strict adherence to API contracts.

\section{Conclusion}
\label{sec:conclusion}

In this paper, we addressed the challenge of hallucination detection in LLM-based API migration. 
We define \scahall as a failure mode where the LLM hallucinates the code required to build the context for calling the target API and use its results.
We identified weaknesses in commonly used similarity metrics like codeBLEU for validating API migrations, showing that they frequently assign high similarity scores to incorrect snippets, when comparing with the ground truth.
We proposed the \hinspector, a lightweight static analysis framework that uses API documentation to check generated code for hallucination. 
Our evaluation demonstrated that this approach significantly outperforms state-of-the-art LLM models acting as judges, reducing false alarms. 
We plan to extend this work by enhancing the API oracle and argument type checking to improve recall of \hinspector.

\section{Acknowledgments}
This work was sponsored by the Engineering and Physical Sciences Research Council (EPSRC) EP/W015927/2.

\balance{}
\bibliographystyle{ACM-Reference-Format}
\bibliography{refs.bib}

\end{document}